\documentclass[%
 reprint,
 amsmath,amssymb,
 aps,
]{revtex4-1}

\usepackage{graphicx}
\usepackage{dcolumn}
\usepackage{bm}

\begin{document}

\preprint{APS/123-QED}

\title{Cosmological Consequences of a Variable Cosmological Constant Model}

\author{Hemza Azri}
 
 \email{hemzaazri@iyte.edu.tr}
\affiliation{%
 Department of Physics, \.{I}zmir Institute of Technology, TR35430
\ \.{I}zmir, Turkey}%

\author{A. Bounames}

\email{bounames@univ-jijel.dz}
\affiliation{%
 University of Jijel, Laboratory of Theoretical Physics, Department of Physics, BP 98, Ouled Aissa, 18000 Jijel, Algeria}%

\date{\today}

\begin{abstract}
We derive a model of dark energy which evolves with time via the scale factor.
The equation of state $\omega=(1-2\alpha)/(1+2\alpha)$ is studied as a function of a parameter $\alpha$
introduced in this model. In addition to
the recent accelerated expansion, the model predicts another decelerated
phase. The age of the
universe is found to be almost consistent with observation. In the limiting
case, the cosmological constant model, we find that vacuum energy gravitates
with a gravitational strength, different than Newton's constant. This enables degravitation of the vacuum energy
which in turn produces the tiny observed curvature, rather than a 120 orders
of magnitude larger value. 
  
\begin{description}

\item[Keywords]
Dark energy, cosmological constant, vacuum energy, accelerating universe.

\item[PACS numbers]
04.20.-q, 95.36.+x, 98.80.-k 
\end{description}

\end{abstract}

\pacs{Valid PACS appear here}
\maketitle


\section{Introduction}

One of the most challenging issue in the history of physics is the implication
of the supernovae data that the cosmic expansion is now accelerating
\cite{Riess1,Perlmutter,Knop,Riess2}. Since then, the simple model of the
cosmological expansion became insufficient, and efforts have been made for a
deep understanding of the origin of this cosmic acceleration by trying to
connect theory with observation. Other important observational data, is the
evidence from the measurement of the cosmic microwave background (CMB) that
besides matter, more than $68\%$ of the energy density of the universe is dark
energy \cite{Planck}. \newline The simplest and well known explanation is the
need to invoke a cosmological constant, a very small and constant amount of
energy with enough negative pressure which acts as repulsive force and causes
the accelerated expansion. Although this appears to explain the cosmic
acceleration, it suffers by the problem of reconciliation of the observed and
theoretical energy density, this was called the cosmological constant problem
(CCP) \cite{Weinberg,Carrol}. This has opened the window to dynamical dark energy
and possible variation of the cosmological constant like any other fundamental 
constants \cite{Terazawa,Dirac}.

As an attempt to generalize Einstein-Schrodinger theory, a gravitational field
equations with a general perfect fluid have been derived \cite{Clerc}. The
spacetime was considered immersed in a larger eight dimensional space, and
this construction leads to a geometrical origin of the velocity vectors due to
the immersion. This theory was considered to give a geometric origin of matter
(and radiation) in the universe via the obtained perfect fluid. An application
of the above theory to cosmology has been done as an attempt to give a
geometrical origin to the cosmological constant \cite{AB}. A time dependent
cosmological constant is derived in that model where its energy density is
decaying via the scale factor.

In this paper, we apply the field equations obtained in \cite{Clerc} to
cosmology, in the case of symmetric connection. The obtained perfect fluid is
interpreted as variable dark energy rather than matter (or radiation) as in
\cite{Clerc}. This dark energy will be derived to be evolving with time via
the scale factor. The ratio of pressure to energy density, the so called
equation of state parameter $\omega$, is studied as a function of another
parameter $\alpha$ as $\omega=(1-2\alpha)/(1+2\alpha)$ and thus we discuss a
two possible decelerated and accelerated phases of the universe. \newline We
will also study the limiting case of this model which is the conventional
cosmological constant model. We will discuss the possible degravitation of the
vacuum energy which produces the tiny observed curvature.

This paper is organized as follows: In section II, we briefly review the model
given in \cite{Clerc}. In section III, we derive the behavior of the dark
energy via the scale factor and discuss the decelerated and accelerated phases
as well as the age of the universe. In section IV, we discuss the idea of
degravitating the cosmological constant from this model. We give our summary
in section V.

\section{The model}

The spacetime of General Relativity is considered to be plunged into a larger
eight dimensional space which has a hypercomplex structure \cite{Clerc,Crum}.
As a result of this construction, in addition to a general asymmetric
connection, the space became endowed with a new antisymmetric tensor of rank
(2,1), denoted $\Lambda_{\alpha\beta}^{\gamma}$. A general energy momentum
tensor of a perfect fluid is derived when this tensor is proposed to have the
form \cite{Clerc,Crum}
\begin{equation}
\Lambda_{\alpha\beta}^{\gamma}=g^{\gamma\sigma}\epsilon_{\sigma\beta\alpha
\rho}U^{\rho},
\end{equation}
where $\epsilon_{\sigma\beta\alpha\rho}$ is the Levi-Civita tensor and
$U^{\rho}$ is an arbitrary four vector. \newline The field equations are
derived from variational principle applied to the following action
\cite{Clerc}
\begin{equation}
S=\int\sqrt{-g}(R-\Lambda)d^{4}x, \label{V6}%
\end{equation}
where the scalar $\Lambda$ is defined as
\begin{equation}
\Lambda=g^{\alpha\beta}\Lambda_{\alpha\rho}^{\gamma}\Lambda_{\beta\gamma
}^{\rho}=6g^{\mu\nu}U_{\mu}U_{\nu}.
\end{equation}

We should mention that this form of the scalar $\Lambda$ given in (\ref{V6})
is not an add hoc term, in fact that is the structure of the spacetime
manifold which is considered as plunged into an eight dimensional manifold
which leads to the presence of this scalar in the Lagrangian. For more details
about the mathematical structure that leads to the action (\ref{V6}), we refer
the reader to the papers \cite{Clerc,AB}. \newline Following the same steps in
\cite{Clerc}, the vector $U_{\mu}$ can be an arbitrary function of the metric
(or $g=\det g_{\mu\nu}$) and coordinates. We propose the particular form for
this vector $U$ as in \cite{Clerc}
\begin{equation}
U_{\mu}=\left(  -g\right)  ^{-\alpha/2}p_{\mu}, \label{form1}%
\end{equation}
where $\alpha$ (noted $q$ in \cite{Clerc}) is a real parameter, $p_{\mu}$ is a
vector density (not a vector), these two quantities are locally only functions
of coordinates of the manifold and are defined such that $U_{\mu}$ is a
vector. \newline Now variation of the action (\ref{V6}) with respect to the
metric tensor gives the field equations
\begin{equation}
R_{\mu\nu}-\frac{1}{2}g_{\mu\nu}R-6U_{\mu}U_{\nu}+\left(  3-6\alpha\right)
U^{2}g_{\mu\nu}=0,
\end{equation}
with $U^{2}=g^{\mu\nu}U_{\mu}U_{\nu}$. In terms of unitary vectors, one can
put
\begin{equation}
U_{\mu}=\lambda u_{\mu}, \label{unit}%
\end{equation}
where $\lambda$ is real and $g^{\mu\nu}u_{\mu}u_{\nu}=1$, then the field
equations become
\begin{equation}
R_{\mu\nu}-\frac{1}{2}g_{\mu\nu}R-6\lambda^{2}u_{\mu}u_{\nu}+\left(
3-6\alpha\right)  \lambda^{2}g_{\mu\nu}=0. \label{Field Eqs}%
\end{equation}
As we see from these equations, one can always define a geometrical
energy-momentum tensor of a perfect fluid as
\begin{equation}
T_{\mu\nu}=6\lambda^{2}u_{\mu}u_{\nu}-\left(  3-6\alpha\right)  \lambda
^{2}g_{\mu\nu}, \label{emtensor}%
\end{equation}
where we took $8\pi G_{N}=1$. \newline This allows us to define an energy
density and pressure of this perfect fluid as \cite{Clerc}
\begin{equation}
\rho=3\lambda^{2}\left(  1+2\alpha\right)  ,\ \ \ p=3\lambda^{2}\left(
1-2\alpha\right)  . \label{energy pressure}%
\end{equation}
One can also write
\begin{equation}
p=\left(  \frac{1-2\alpha}{1+2\alpha}\right)  \rho. \label{rel}%
\end{equation}
The above model has been derived in \cite{Clerc,Crum} and the quantities found
in equation (\ref{energy pressure}) are interpreted as energy density and
pressure of matter (and radiation). \newline Next, by assuming that matter is
not a geometric quantity, we shall interpret the quantities given in
(\ref{energy pressure}) as the energy density and pressure of dark energy.

\section{Equation of state and age of the universe}

In this section, we will derive the evolution of the dark energy density (as
well as pressure) in terms of the cosmological scale factor $a\left(  t\right)
$. As in standard cosmology we apply the covariant conservation law to the
energy momentum tensor (\ref{emtensor}). In the flat Friedmann Robertson
Walker spacetime, this law (continuity equation) gives
\begin{equation}
\left( 1+2\alpha\right)  \dot{\lambda}+3\frac{\dot{a}}{a}\lambda=0,
\label{diff}%
\end{equation}
which can be solved easily as
\begin{equation}
\lambda=\lambda_{0}\left(  \frac{a}{a_{0}}\right) ^{-\frac{3}{\left(
1+2\alpha\right) }}, \label{lambda}%
\end{equation}
where $a_{0}$ is the scale factor at present and $\lambda_{0}=\lambda\left(
a_{0}\right)  $ is a constant.

Now the energy density given in (\ref{energy pressure}) which we interpret
here as dark energy density, takes the form
\begin{equation}
\rho^{DE}=3\lambda_{0}^{2}\left(  1+2\alpha\right)  \left(  \frac{a}{a_{0}%
}\right) ^{-\frac{6}{\left( 1+2\alpha\right) }}, \label{dens}%
\end{equation}
where one can put $\rho_{0}^{DE}=3\lambda_{0}^{2}\left(  1+2\alpha\right) $,
the Dark Energy density when $a=a_{0}$. The pressure is now given by its
relation with the energy density (\ref{rel}). \newline As we see, this model
describes dark energy which evolves with time via the scale factor, which does
not decay faster than matter $\rho^{M}$ $\sim a^{-3}$, in the recent
accelerated phase given by its conditions on $\alpha$ that we will see later.

\begin{figure}[tbh]
\begin{centering}
\includegraphics[width=3in,height=2in]{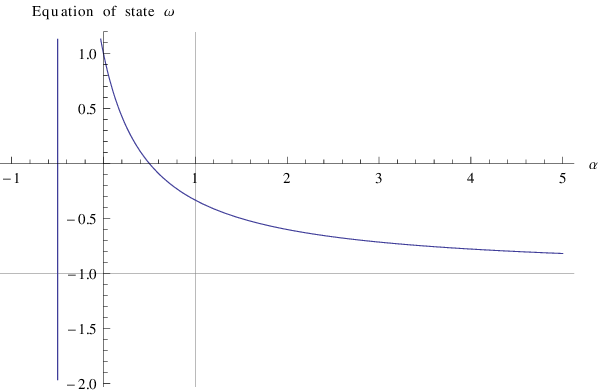}
\par\end{centering}
\caption{The equation of state parameter $\omega$ (vertical axis) as a
function of the parameter $\alpha>-1/2$ (horizontal axis). The model predicts
a decelerated and accelerated phases, corresponding to $\alpha<1$ and
$\alpha>1$ respectively. The standard cosmological constant model, i.e
$\omega=-1$ appears for very large values of $\alpha$.}%
\label{fig:omegade}%
\end{figure}

The ratio of pressure to energy density, i.e; the equation of state parameter
$\omega=p/\rho$ gives a useful description for dark energy. For this model
described by the energy density and pressure given by (\ref{energy pressure}),
the equation of state parameter is written as
\begin{equation}
\omega=\frac{1-2\alpha}{1+2\alpha}. \label{omeg}%
\end{equation}
As we see from this expression, for very big $\alpha$, this model coincides
with the standard cosmological constant constant model where $\omega=-1$.
Here, cosmological constant means a term $\Lambda$ (may not be constant) which
appears in Einstein's equations as $\Lambda g_{\mu\nu}$. Later in this paper,
we will study the case of a strictly cosmological constant obtained here from
the field equations (\ref{Field Eqs}) when $\alpha$ is very large. \newline
From the expression (\ref{omeg}), one can expect different phases depending on
the chosen values of the real parameter $\alpha$. The equation of state
parameter (\ref{omeg}) is plotted in Figure \ref{fig:omegade} for some values
of $\alpha$.

As one can see from the figure, the accelerating phase $\omega<-1/3$ is for
$\alpha>1$. The case $-1/2<\alpha<1$ corresponds to a decelerating phase
$\omega>-1/3$, while acceleration vanishes, i.e; $\omega=-1/3$ for $\alpha=1$.
\newline We should mention that this study is made for a universe dominated by
only dark energy. In standard cosmology, the accelerated and decelerated
universes correspond to some specific relations between matter and dark
energy. In fact, for the simple model of dark energy described by a
cosmological constant, the decelerated phase corresponds to the era in which
the matter energy density $\rho^{M}$ was bigger than about $2\rho^{DE}$, and
the universe started accelerating when $\rho^{M}$ became less than $2\rho
^{DE}$. \newline In our model, the above relations between matter and dark
energy densities in the decelerated and accelerated eras are different from
the standard cosmological constant model. In fact, in the presence of both
matter and dark energy, one can write the second Friedmann equation in this
case as
\begin{equation}
\left(  \frac{\overset{\cdot\cdot}{a}}{a}\right)  =-\frac{4\pi G_{N}}%
{3}\left[  \rho^{M}+\frac{4\left(  1-\alpha\right)  }{1+2\alpha}\rho
^{DE}\right]  , \label{Fried2}%
\end{equation}
where we have used relation (\ref{rel}) for dark energy. \newline As we see
from (\ref{Fried2}), for very large values of $\alpha$, the second term in the
right hand side becomes $-2\rho^{DE}$, and then we get the cosmological
constant model discussed above. \newline In contrast to the standard
cosmological constant model, as we see from equation (\ref{Fried2}), the
decelerated and accelerated phases correspond respectively to matter energy
density greater and less than the quantity $\frac{4\left(  \alpha-1\right)
}{1+2\alpha}\rho^{DE}$, rather than $2\rho^{DE}$. In the case $\alpha=1$, only
matter energy density appears in the right hand side of (\ref{Fried2}), which
means that this case corresponds to a decelerated phase of the universe, this
is not in contradiction with the previous study where acceleration vanishes
for this case, because we have taken a universe dominated by only dark energy
while here, matter is included. \newline In most of dark energy models one
considers only the accelerated phase. Here in our model, the decelerated phase
has its origin from the form of the energy momentum tensor (\ref{emtensor})
where its first term behaves as a matter term.

As we have seen so far, the phases of the universe are studied in this model
as a function of the real parameter $\alpha$ which unfortunately the model
does not offer a mechanism to fix it. This opens the problem of the nature of
this parameter and its dependence on the physical parameters for instance time
and temperature which are the best physical parameters that can be used to
study the epochs of the universe.

In the simplest model for a spatially flat universe filled with only matter,
the age of the universe is $t_{0}=\frac{2}{3}H_{0}^{-1}$ billion years which
is found to be shorter than the ages of some oldest stars in globular
clusters, $12\lesssim t_{0}\lesssim15$ billion years \cite{Krauss}. This age
problem is solved by introducing the conventional cosmological constant to be
$t_{0}\sim H_{0}^{-1}$ billion years \cite{Planck}. \newline\newline As we
shall see later, the model given here predicts the same age as in the
cosmological constant model where the universe is supposed to be dominated by
dark energy. The solution of the first Friedman equation in this model is
\begin{equation}
a\left(  t\right)  =a_{0}\left[  1+\frac{3}{1+2\alpha} \sqrt{\frac{8\pi G_{N}%
}{3}\rho_{0}^{DE}}\left(  t-t_{0}\right)  \right] ^{\frac{1+2\alpha}{3}},
\end{equation}
where we have used the dark energy density
\begin{equation}
\rho^{DE}=\rho_{0}^{DE}\left( a/a_{0}\right) ^{-\frac{6}{1+2\alpha}}.
\end{equation}

If we include matter in addition to dark energy, one may easily solve the
first Friedman equation, and obtain the age of the universe in terms of the
parameters $\Omega_{m}$, $H_{0}$ and $\alpha$ as follows
\begin{equation}
t_{0}=H_{0}^{-1}\int_{0}^{1}\frac{dx}{x\sqrt{\Omega_{m}x^{-3}+(1-\Omega
_{m})x^{-\frac{6}{1+2\alpha}}}}%
\end{equation}

For the accelerating case, i.e, the range $1\leq\alpha\leq\infty$, the last
relation gives an age for the universe $0.80 H^{-1}_{0} \leq t_{0} \leq1.12
H^{-1}_{0}$, for $\Omega_{m}\simeq0.31$. Recent results estimate the age of
the universe to be $13.7$ billion years \cite{Planck}, which is in the range
obtained here.

We conclude this section by stating that models of a varying cosmological
constant similar to our model have been considered by some authors. In most of
these models, the form of the cosmological constant was proposed (ad hoc) for
some cosmological reasons \cite{Chen,Lopez}, while in this paper, we derived
it from geometry.

\section{Gravitational constant and the cosmological constant problem}

The model studied above is a general framework of variable dark energy which
allows us to study different phases of the universe depending on the free
parameter $\alpha$ which determines the equation of state $\omega$. \newline
As we have mentioned during this study, a cosmological constant term can be
obtained as a limit of our model for a very large $\alpha$. In fact, in this
case the field equations (\ref{Field Eqs}) become
\begin{equation}
R_{\mu\nu}-\frac{1}{2}g_{\mu\nu}R\simeq6\alpha\lambda^{2}g_{\mu\nu}.
\end{equation}
In terms of strictly 'constant' cosmological constant, these equations can be
written as
\begin{equation}
R_{\mu\nu}-\frac{1}{2}g_{\mu\nu}R  =\lambda^{2}\Lambda
g_{\mu\nu},
\end{equation}
where we have defined a cosmological constant $\Lambda=6\alpha$ (independent
from the scalar $\lambda$). In terms of the vacuum energy density
$\rho^{vac}=\Lambda M_{Pl}^{2}$, where $M_{Pl}=\left(  8\pi G_{N}\right)
^{-1/2}$ is the Planck Mass, the last equation is written as
\begin{equation}
R_{\mu\nu}-\frac{1}{2}g_{\mu\nu}R=\left(  \frac{M_{Pl}}{\lambda}\right)
^{-2}\rho^{vac}g_{\mu\nu}.\label{vacuum}%
\end{equation}
In this case, we have $p\simeq -\rho$ (cosmological constant), and the continuity equation is solved as $\lambda=\lambda_{0}$ (a constant) rather than solution (\ref{lambda}). Nevertheless, this is also clear from the solution (\ref{lambda}) when $\alpha$ is large enough.

In general relativity, ordinary matter and vacuum energy gravitate with Newton's constant. For the material part, this is supported and confirmed by experiments and observations, both at the solar system and the cosmological scales.

However, vacuum energy, originated from zero point energies of quantum fields as well as phase transitions, makes a perplexed problem when seen in the framework of general relativity. The zero-point energies of quantum fields are of the order $\Lambda_{\text{UV}}^{4}$, where $\Lambda_{\text{UV}}$ is the
Ultra-Violet Cutoff. The spacetime curvature is very sensible to this quantity, in fact, if we trust quantum field theory up to Planck scale, i.e, $\Lambda_{\text{UV}}=M_{Pl}$, the scalar curvature that corresponds to this vacuum is $R^{\text{theo}}\sim M_{Pl}^{2}$. This theoretical value, estimated from ground states of particle fields, severely contradicts the observed curvature $R^{\text{obs}}\sim 10^{-47}\text{eV}^{2}$. This discrepancy, which is about $120$ orders of magnitude between theory and observation is the origin of the cosmological constant problem.

In terms of mass scales, and for no physical reason, the related observed vacuum energy density is of the order of the Neutrino mass density, and the mentioned scalar curvature is given as 
\begin{equation}
\label{obs}
R^{\text{obs}}\sim\frac{m_{\nu}^{4}}{M_{Pl}^{2}}, 
\end{equation}
where $m_{\nu}\simeq 10^{-3}\text{eV}$ is the Neutrino mass.

Although the theoretical value of the vacuum energy is estimated from quantum field theory, the cosmological constant problem resides essentially in general relativity where this vacuum gravitates with Newton's constant. In the present model and for this limit described by the field equation (\ref{vacuum}), this requirement imposes with no physical reason, $\lambda_{0}=1$. However, it is only observational bounds on the curvature $R^{\text{obs}}$ that determine the correct value of this constant. A value $\lambda_{0} \neq 1$ translates the De-gravitation of the cosmological constant; unlike ordinary matter, vacuum does not gravitate with Newton's constant. 

Dirac large number hypothesis implies reasonable form of the constant $\lambda_{0}$ due to the very tiny observed curvature. This form might be a ratio of two hierarchically mass scales  \cite{Dirac}. At that end, this ratio can be written as
\begin{equation}
\label{lambda0}
\lambda_{0}=\frac{M_{Pl}}{M_{Co}},
\end{equation}
where we have proposed the new cosmological strength $M_{Co}$ in addition to the fundamental gravity mass scale $M_{Pl}$.

Thus, the observational bound on the curvature (\ref{obs}) and the form of this later in terms of the theoretical cosmological constant $\Lambda$ lead to the following constraint on $M_{Co}$ \cite{Demir,Azri}
\begin{equation}
\label{mcos}
M_{Co}^{2}\simeq \Lambda \left(\frac{M_{Pl}}{m_{\nu}} \right)^{4}. 
\end{equation}
This shows clearly that the cosmological constant gravitates with $M_{Co}$, where for the larger value $\Lambda=M_{Pl}^{2}$, this mass scale is the mass of the universe \cite{Demir,Nima}. It is this case that fixes correctly the constant $\lambda_{0}$, in fact, from equations (\ref{lambda0}) and (\ref{mcos}) we get
\begin{equation}
\lambda_{0}=\left(\frac{m_{\nu}}{M_{Pl}} \right)^{2}.
\end{equation}  

In the recent years, degravitating the cosmological constant became of great
interest. Unlike ordinary matter which gravitates with the Newton's
gravitational constant, vacuum energy is considered to gravitate with
different cosmological strengths \cite{Demir,Azri,Nima}. Although they lack a real
mechanism to derive the new cosmological strengths, these models (including
the one given in this paper) enable degravitation of the cosmological
constant, such that vacuum energy produces the observed tiny curvature due to
hierarchy between different mass scales rather than fine-tuning.

\section{Summary}

In this paper, we have studied a model with variable dark energy. This model
is derived from a geometrical construction used by some authors as an attempts
to generalize the unified theory of Einstein-Schrodinger. Unlike the
interpretation given in that work, we have interpreted here the obtained
energy momentum tensor as dark energy. \newline We have studied some
cosmological consequences of this model where dark energy is found to be
evolving with time via the scale factor. Unlike most of dark energy models
where the studies are restricted only to the accelerated expansion, here the
model predicts another decelerated phase of the universe. The reason is an
additional term that appears in the energy momentum tensor, in addition to the
cosmological term. The equation of state $\omega$ is found to depend on a real
parameter $\alpha$ and the condition of accelerating and decelerating phases
has been studied as a function of this parameter. The model does not offer a
mechanism to fix this parameter which we believe should depend on some of the
physical parameters such as time and temperature. \newline We have studied the
limiting case which is the cosmological constant model. We found that vacuum
energy can be degravitated, in other word unlike matter and radiation which
gravitate with Newton's constant, vacuum energy in this model gravitates with
a different gravitational strength which is fixed here using observational
considerations. As a result, vacuum energy which receives its value from
zero-point energies of quantum fields as well as early time phase transitions,
curves the empty space by a tiny amount consistent with the observed value.
\newline 

\begin{acknowledgments}
The author H. A. would like to thank Durmu{\c{s}} Ali Demir for useful
comments and Ren\'{e}-Louis Clerc for correspondences about the mathematical structure given in Sec.II.
\end{acknowledgments}

\end{document}